# Nanostructure analysis of InGaN/GaN quantum wells based on semi-polar-faced GaN nanorods

Yu-Sheng Huang,[1] Yu-Hsin Weng,[1] Yung-Sheng Chen,[1] Shih-Wei Feng,[2,6] Chie-Tong Kuo,[3] Ming-Yen Lu,[4] Yung-Chen Cheng,[5] Ya-Ping Hsieh,[1] and Hsiang-Chen Wang,[1,*]

[1] Graduate Institute of Opto-Mechatronics, National Chung Cheng University, 168 University Rd., Min-Hsiung, Chia-Yi 62102, Taiwan
[2] Department of Applied Physics, National University of Kaohsiung, 700 Kaohsiung University Rd., Nanzih District, Kaohsiung 81148, Taiwan.
[3] Department of Physics, National Sun Yat-sen University, 70 Lienhai Rd., Kaohsiung 80424, Taiwan.
[4] Department of Materials Science and Engineering, National Tsing Hua University, 101, Sec. 2, Kuang-Fu Road, Hsinchu 30013, Taiwan
[5] Department of Materials Science, National University of Tainan, Tainan 70005, Taiwan
[6]swfeng@nuk.edu.tw
*hcwang@ccu.edu.tw

**Abstract**: We demonstrate a series of InGaN/GaN double quantum well nanostructure elements. We grow a layer of 2 μm undoped GaN template on top of a (0001)-direction sapphire substrate. A 100 nm $SiO_2$ thin film is deposited on top as a masking pattern layer. This layer is then covered with a 300 nm aluminum layer as the anodic aluminum oxide (AAO) hole pattern layer. After oxalic acid etching, we transfer the hole pattern from the AAO layer to the $SiO_2$ layer by reactive ion etching. Lastly, we utilize metal-organic chemical vapor deposition to grow GaN nanorods approximately 1.5 μm in size. We then grow two layers of InGaN/GaN double quantum wells on the semi-polar face of the GaN nanorod substrate under different temperatures. We then study the characteristics of the InGaN/GaN quantum wells formed on the semi-polar faces of GaN nanorods. We report the following findings from our study: first, using $SiO_2$ with repeating hole pattern, we are able to grow high-quality GaN nanorods with diameters of approximately 80-120 nm; second, photoluminescence (PL) measurements enable us to identify Fabry-Perot effect from InGaN/GaN quantum wells on the semi-polar face. We calculate the quantum wells' cavity thickness with obtained PL measurements. Lastly, high resolution TEM images allow us to study the lattice structure characteristics of InGaN/GaN quantum wells on GaN nanorod and identify the existence of threading dislocations in the lattice structure that affects the GaN nanorod's growth mechanism.

## References and links

## 1. Introduction

Ever since Shuji Nakamura first produced GaN thin films via metal-organic chemical vapor deposition (MOCVD) in 1993, thin films have aroused great interest and various research efforts in the academic field [1]. Given that IIIA group-N materials and their alloys possess direct energy gaps, which emit light over the entire visible light range [2], these materials have potential use in next-generation lighting applications [3]. Nano-GaN materials possess unique physical characteristics attributed to their low-dimensional effects, such as lower melting point and different energy gaps [4]. Nano-GaN materials are utilized as various electronic elements because of their broader light emission bands (from ultraviolet to infrared) and their ability to continue functioning under high temperature and pressure conditions [5]. To improve the quality of InGaN epitaxy grown on sapphire substrate and to solve lattice structure mismatch between the sapphire substrate and GaN, researchers have developed epitaxy techniques, such as epitaxial lateral overgrow [6], micro-level SiNx with patterned SiNx interlayer structures, and patterned sapphire substrates [7,8]. Research efforts related to growing GaN on non-polarized sapphire substrates have gained considerable attention in recent years, especially because of the results reported by S. Nakamura and his team [9–11]. Other than growth on non-polar a-, m-, and r- planes of three sapphire substrate lattice planes, growth on semi-polar (11–22) planes of sapphire substrates have recently been developed. Semi-polar (11–22) planes of sapphire substrates have the advantage of less lattice structure mismatches with GaN thin films as well as more moderate quantum well barriers.

In our study, we grow bonded GaN nanorods with a $SiO_2$ masking hole pattern to reduce threading dislocations (TDs) caused by lattice structure mismatch between the sapphire substrate and GaN to obtain GaN nanorods with better quality [12,13]. We then utilize hexagonal structure behavior (nanorod growth speed on C-axis (0001) direction is much higher than that of semi-polar (11–22) face) to obtain GaN nanorods with six semi-polar faces on the top. We then study the optical and material properties of InGaN/GaN quantum wells grown on semi-polar faces.

## 2. Growth conditions and sample structure

We use MOCVD to grow a series of GaN nanorod substrates. The process is shown in Figure 1. First, a layer of 2 μm undoped GaN template is placed on top of the sapphire substrate (Fig. 1(a)). A 100 nm $SiO_2$ thin film is deposited by e-beam on the template as the masking pattern (Fig. 1(b)). A 300 nm aluminum layer is sputtered on as the AAO hole pattern layer (Fig. 1(c)). We then conduct secondary oxalic acid etching (0.3 M oxalic aide with 40 V) on this layer to form the hole pattern (Fig. 1(d)). The hole pattern is then transferred to $SiO_2$ thin film by reactive ion etching (Fig. 1(e)). Afterward, the AAO layer is removed with a mixture of 6% phosphoric acid and 1.5% chromic acid under 333 °C (Fig. 1(f)). Lastly, we apply MOCVD under 1020 °C for 1 min to grow nanorods (Fig. 1(g)).

This study compares four different samples under different growing conditions. Sample A is the reference sample and is a GaN nanorod grown on $SiO_2$ hole pattern without any quantum wells. Samples B-D are grown on two layers of InGaN/GaN quantum wells on GaN nanorods. Indium has a concentration of 5%. The quantum wells have widths of 2-2.5 nm and barrier heights of 10-12 nm. The quantum wells are grown under 750, 700, and 650 °C, whereas the barrier is formed under 870 °C. The structural schematic is shown in Figure 1(h). Given that our previous study has identified the correlation between the growing temperature of quantum wells and the growth of InGaN, the present study attempts to control the growing temperature and study the differences among InGaN/GaN quantum wells. The structures of grown InGaN/GaN quantum wells are shown in Figure 1 (f).

We use photoluminescence (PL), scanning electron microscopy (SEM), and transmission electron microscopy (TEM) to study the samples produced by the method mentioned above. PL spectra are obtained with the 325 nm line of a 50 mW He-Cd laser as the excitation source. Samples are placed in a cryostat for temperature-dependent measurements. Temperature range is 10–290 K. SEM observations are performed with a JEOL JSM 6700F system. Finally, TEM observations are performed with a JEOL TEM-300F field emission electron microscope with an acceleration voltage of 300 KV.

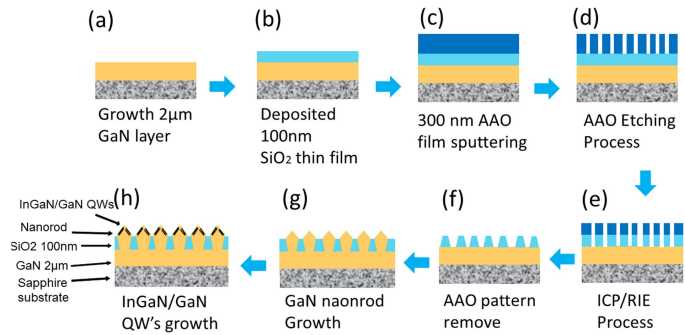

Fig 1. Fabrication process of the s semi-polar faced InGaN/GaN double quantum wells nanorods samples.

## 3. PL measurements

Figures 2 (a)-(d) show the PL measurement results for samples A-D. We excite the sample with a 325 nm laser under 10-290 K. Figure 2 (a) shows a peak at approximately 355-360 nm, which is the peak position of GaN. The high peak intensity and narrow peak profile confirm that our procedure produced high-quality bonded GaN nanorod structures [14-17]. Figures 2 (b)-(d) show the PL measurements of InGaN/GaN double quantum wells under growth temperatures of 750, 700, and 650 °C. The GaN nanorod samples have constant peak positions at 355-360 nm and similar peak intensities but a rather wider peak profiles as sample A. The width of the peak profile increases as the temperature decreases, indicating a less uniformed GaN structure. We also discover a peak at 378-400 nm, which corresponds to the peak position for InGaN quantum wells with 5% concentration. Previous studies have reported that these peaks will not change with temperature [18]. Such peaks show an increase in intensity as well as slightly redshift as temperature decreases. This behavior may cause by bandgap shrinkage as temperature decreases. Closer examination shows that there are two peaks in 378-400 nm. These peaks represent two slightly different concentrations of InGaN quantum wells. We believe that this concentration difference is caused by lattice structure mismatches. The formation of the first layer of the quantum well (bottom layer) reduces lattice structure mismatch, thus resulting in higher InGaN concentrations in the second layer (top layer). This behavior is verified by the slightly lower intensity of the longer wavelength. The second peak represents a second layer with higher concentration. Figure 2 (d) shows the PL measurement for sample D. Here, we see a far more complicated peak pattern at approximately 375-425 nm. Along with the two peaks seen in samples B and C, there are two smaller peaks at approximately 400 nm and 418 nm [19]. These peak distributions resemble Fabry-Perot resonance modes [20]. This behavior is caused by the resonance cavity formed by the InGaN/GaN quantum well structures. To obtain the width of the quantum wells, we apply these peak positions to the following formula:

$$\frac{i}{d} = \frac{2n(\lambda)}{\lambda_{peak}} \qquad (1)$$

Where i is the peak number, n(λ) is the index for GaN as a function of wavelength, $\lambda_{peak}$ is the peak wavelength, and d is the width of the quantum well.

We obtain d as 2.4 nm, which corresponds with the InGaN quantum well width obtained by our previous study. This result validates the quality of InGaN quantum wells grown on GaN nanorods and the application of Fabry-Perot effect on perfectly grown InGaN/GaN quantum wells. Moreover, the extra peaks in sample D imply more resonance cavity, which is beneficial in illuminating device applications. From the PL measurement results, we identify that a lower growth temperature results in better grown InGaN/GaN quantum wells at a cost of less uniformed GaN structure. We also verify the formation of quantum wells by Fabry-Perot resonance modes. However, there is still much to understand about the structural characteristics of quantum wells. Thus, we further analyze our samples and their illumination amplifying effects with SEM and TEM imaging.

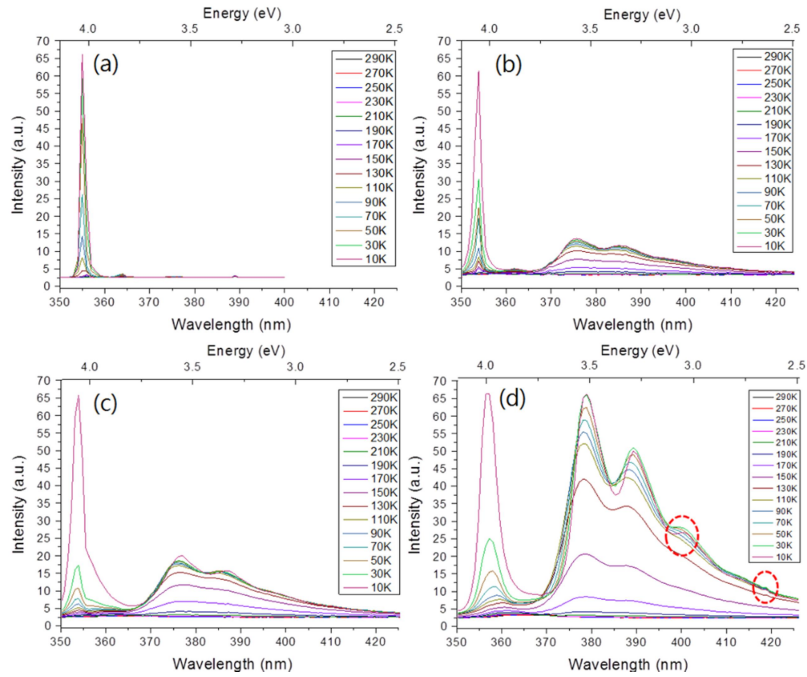

Fig 2. PL results of the semi-polar-faced InGaN/GaN double quantum well nanorods. (a) Sample A. (b) Sample B. (c) Sample C. (d) Sample D.

## 4. SEM measurements

We perform SEM to obtain the top and cross-section views of samples A (Figures 3 (a), (b)) and D (Figure 3s (c), (d)). The cross-section view of sample A (Figure 3 (b)) shows that the grown nanorod is approximately 100-150 nm tall and has a 6-faced hexagonal structure. The cross-section view of sample D (Figure 3(d)) shows a mushroomed-shaped GaN nanorod capped with an InGaN/GaN quantum well. Its total height is approximately 120-170 nm. Its cap diameter is approximately 130-150 nm. We believe that this cap structure is formed by the outward extension of the quantum well layer on top of the nanorod. The top view of sample A (Figure 3(a)) shows empty regions on the $SiO_2$ substrate alongside the grown GaN nanorods. The formation of these empty regions requires further investigation and is discussed in the next section. On the other hand, the top view of Sample D (Figure 3(c)) shows closer-spaced GaN nanorods. Reduced spacing can also be explained by the outward extension of the quantum well layer.

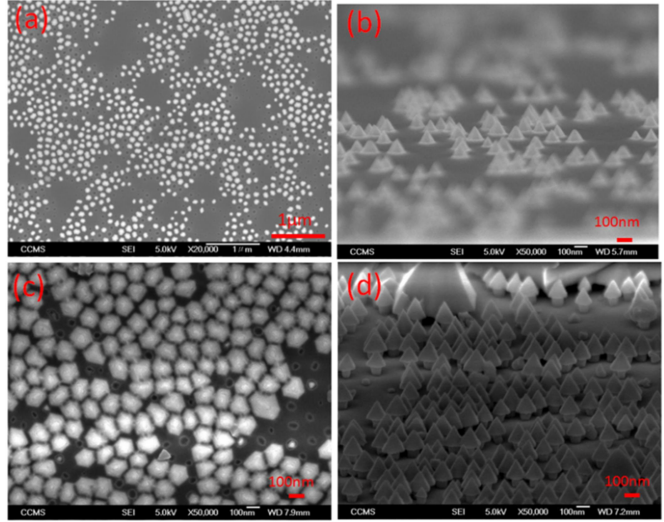

Fig 3. SEM images of the semi-polar-faced InGaN/GaN double quantum well nanorods. (a) Top view of sample A. (b) Cross-section view of sample A. (c) Top view of sample D. (d) Cross-section view of sample D.

## 5. TEM and HRTEM measurements

Figure 4 shows the cross-sectional TEM image for samples A to D. We deduce the growing process of GaN nanorods from Figure 4 (a): GaN first grows from the GaN film layer, which is bonded by the hole pattern on the $SiO_2$ layer. Afterward, the GaN nanorod grows over the $SiO_2$ restriction and begins forming a hexagonal, 6-faced structure. This hexagonal structure results from the difference in growth speed among the GaN lattice structures. Lattice structures grow considerably faster in the (0001) direction than in other semi-polar (11-22) faces. From Figure 4 (b)-(d), we also observe that InGaN/GaN quantum wells extend outside the nanorods to form a cap structure along the $SiO_2$ layer. These cap structures can even contact other caps. We further investigate the nanorod structures by high-resolution TEM (HRTEM). In Figure 5 (a), the red circle indicates the junction between the GaN film and $SiO_2$ layer; thus, we can verify if the GaN nanorod and GaN film layer are perfectly integrated into a uniform structure. The lattice structure between two layers is continuous and there is no discontinuity in the (0001) direction. Figure 5 (b) includes areas without grown GaN nanorods. We observe a slightly different lattice structure in the GaN film layer, as indicated by the blue circle. We believe these areas are caused by threading dislocations (TDs) between the GaN film and SiO$_2$ layer [21, 22]. A massive, V-shaped hole on top of the TDs prevents the attachment of GaN crystals during the epitaxial process, resulting in empty bases on the GaN layer [23].

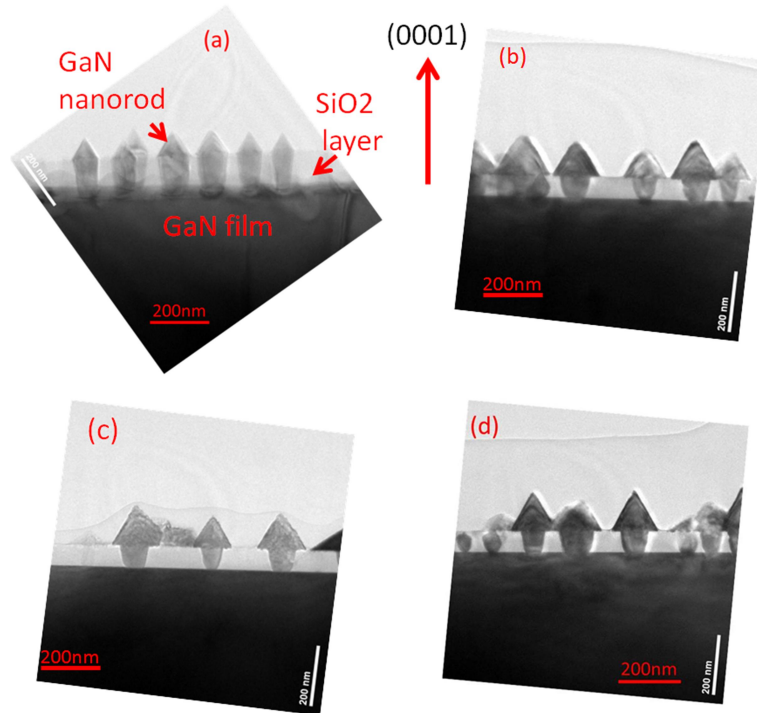

Fig 4. TEM images of the semi-polar-faced InGaN/GaN double quantum well nanorods. (a) Sample A. (b) Sample B. (c) Sample C. (d) Sample D.

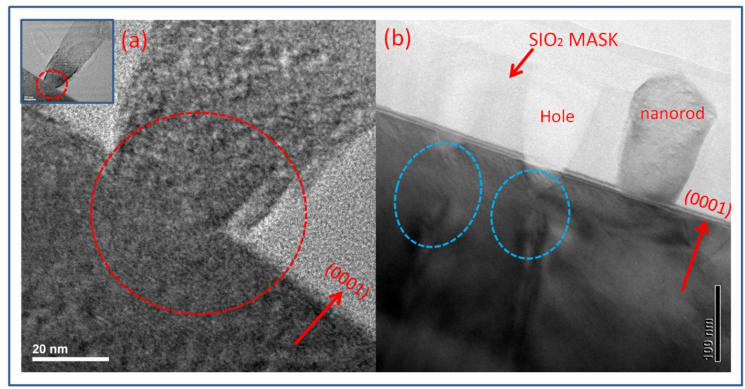

Fig 5. HRTEM images of the GaN and SiO$_2$ layers of the produced GaN nanorods. (a) Close-up view of the atomic lattice image of the GaN nanorod base. (b) Zoomed-out view of the atomic lattice image of the GaN nanorod base.

Figures 6 (a) and (b) respectively show the HRTEM images of the left side and right side of InGaN/GaN quantum well structure (cap structure) on top of the nanorod, as indicated by the small two red circles. The black stripes are GaN lattice structure layers grown in the (0001) direction. Given that InGaN has a larger mass number than GaN, electron beams from TEM will have a larger scattering angle, thus resulting in the darker regions in the figure. The two darker lines running almost parallel to the cap structure's edges are the two InGaN/GaN quantum wells. The InGaN quantum well is approximately 2.5 nm wide and corresponds with our Fabry-Perot effect calculation; the GaN barrier is approximately 10 nm [24-26]. The angle between the quantum well structures and GaN lattice direction is approximately 58.45°. This angle indicates that the quantum wells are grown from the semi-polar (11-22) GaN lattice face. Another possible growth direction is the semi-polar (1-101) GaN lattice face, which is approximately 61.9° to the (0001) direction; however, no angle greater than 60° is observed [27,28]. We confirm that the quantum wells are grown in the semi-polar (11-22) face direction. Quantum wells on the semi-polar (11-22) face less lattice mismatches between InGaN and GaN, as reported by previous studies [29-36]. We believe that less lattice mismatches result in better-quality GaN nanorods and InGaN/GaN quantum wells. Thus, they are more closely spaced, as seen in Figure 2.

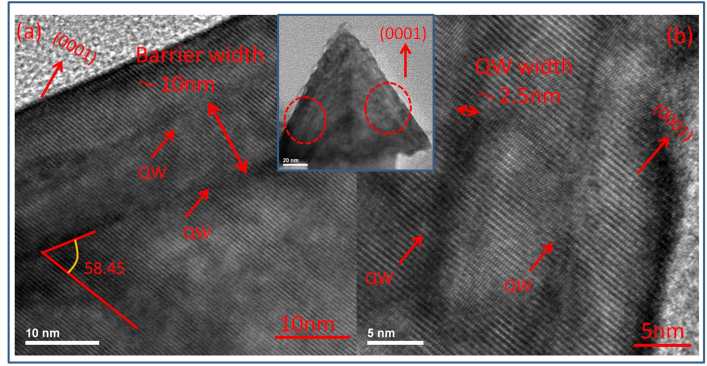

Fig 6. HRTEM images of the semi-polar-faced InGaN/GaN double quantum well cap structure. (a) Structural atomic lattice image of sample D's left-side cap. (b) Structural atomic lattice image of sample D's right-side cap.

## 6. Conclusion

We grow a series of GaN nanorod structures with InGaN/GaN double quantum wells on semi-polar (11–22) faces. We study the samples' photoluminescence peak profiles and structures with PL measurements. We verify the good quality of our GaN nanorods. We also confirm that lower growth temperatures result in higher quality InGaN/GaN double quantum wells. We observe the Fabry-Perot effect in our PL measurements; this effect results from the resonance cavity caused by the formation of InGaN/GaN double quantum wells. Sample D presents two extra resonance cavities that may be useful in amplifying illumination. From SEM images, we observe the overall configuration and the cap-like structures caused by the extension of quantum wells. These structures are further investigated by TEM and HRTEM. On the GaN layer, we identify the TDs affecting GaN nanorod growth. HRTEM allows us to closely study the quantum well structure and verify that InGaN/GaN double quantum wells grow in the semi-polar (11–22) face direction. These GaN nanorods form a hexagonal structure on top of the GaN nanorod with 6 semi-polar (11–22) faces.

In conclusion, we believe that our method, which utilizes a bonded $SiO_2$ masked layer to reduce lattice structure mismatch, produces high-quality GaN nanorods with InGaN/GaN double quantum wells on semi-polar (11–22) faces. We also identify the structural characteristic of the obtained high-quality GaN nanorods with InGaN/GaN double quantum wells.